\documentclass[letter,superscriptaddress,twocolumn,prl,showkeys,showpacs]{revtex4-1}

\usepackage[utf8]{inputenc}
\usepackage{amsmath}   
\usepackage{graphicx}   
\usepackage{verbatim}  
\usepackage{color}     
\usepackage{subfigure}  
\usepackage{hyperref}   
\usepackage{natbib}
\usepackage{gensymb }
\usepackage{color}
\usepackage{epstopdf}
\usepackage{multirow}

\begin{document}

\title{Short-range phase coherence and origin of the $1T$-TiSe$_2$ charge density wave}

\author{B. Hildebrand}
\altaffiliation{Corresponding author.\\ baptiste.hildebrand@unifr.ch}
\affiliation{D{\'e}partement de Physique and Fribourg Center for Nanomaterials, Universit{\'e} de Fribourg, CH-1700 Fribourg, Switzerland}

\author{T. Jaouen}
\altaffiliation{Corresponding author.\\ thomas.jaouen@unifr.ch}
\affiliation{D{\'e}partement de Physique and Fribourg Center for Nanomaterials, Universit{\'e} de Fribourg, CH-1700 Fribourg, Switzerland}

\author{C. Didiot}
\affiliation{D{\'e}partement de Physique and Fribourg Center for Nanomaterials, Universit{\'e} de Fribourg, CH-1700 Fribourg, Switzerland}

\author{E. Razzoli}
\affiliation{D{\'e}partement de Physique and Fribourg Center for Nanomaterials, Universit{\'e} de Fribourg, CH-1700 Fribourg, Switzerland}

\author{G. Monney}
\affiliation{D{\'e}partement de Physique and Fribourg Center for Nanomaterials, Universit{\'e} de Fribourg, CH-1700 Fribourg, Switzerland}

\author{M.-L. Mottas}
\affiliation{D{\'e}partement de Physique and Fribourg Center for Nanomaterials, Universit{\'e} de Fribourg, CH-1700 Fribourg, Switzerland}

\author{A. Ubaldini}
\affiliation{Department of Quantum Matter Physics, University of Geneva, 24 Quai Ernest-Ansermet, 1211 Geneva 4, Switzerland}

\author{H. Berger}
\affiliation{Institut de G{\'e}nie Atomique, Ecole Polytechnique F{\'e}d{\'e}rale de Lausanne, CH-1015 Lausanne, Switzerland}

\author{C. Barreteau}
\affiliation{Department of Quantum Matter Physics, University of Geneva, 24 Quai Ernest-Ansermet, 1211 Geneva 4, Switzerland}

\author{H. Beck}
\affiliation{D{\'e}partement de Physique and Fribourg Center for Nanomaterials, Universit{\'e} de Fribourg, CH-1700 Fribourg, Switzerland}

\author{D. R. Bowler}
\affiliation{London Centre for Nanotechnology and Department of Physics and Astronomy, University College London, London WC1E 6BT, UK}

\author{P. Aebi}
\affiliation{D{\'e}partement de Physique and Fribourg Center for Nanomaterials, Universit{\'e} de Fribourg, CH-1700 Fribourg, Switzerland}

\begin{abstract}
The impact of variable Ti self-doping on the $1T$-TiSe$_2$ charge density wave (CDW) is studied by scanning tunneling microscopy. Supported by density functional theory we show that agglomeration of intercalated-Ti atoms acts as preferential nucleation centers for the CDW that breaks up in phase-shifted CDW domains whose size directly depends on the intercalated-Ti concentration and which are separated by atomically-sharp phase boundaries. The close relationship between the diminution of the CDW domain size and the disappearance of the anomalous peak in the temperature dependent resistivity allows to draw a coherent picture of the $1T$-TiSe$_2$ CDW phase transition and its relation to excitons. 
\end{abstract}
\date{\today}
\maketitle

The understanding of broken-symmetry ground states is facing the complexity of the many-body problem arising from the multiplicity of coupled electron-electron, electron-hole, and electron-lattice interactions. A typical material exhibiting this close entanglement is the quasi-two dimensional, layered transition-metal dichalcogenide $1T$-TiSe$_2$ consisting of hexagonal 1$T$-stacked Se-Ti-Se layers separated by a van der Waals (vdW) gap. At $T_{\text{CDW}}$ $\sim$200 K, it undergoes a phase transition towards a 2x2x2 commensurate CDW phase at which a weak periodic lattice distortion (PLD) develops and it shows  a broad maximum in the temperature dependent resistivity in the vicinity of  $T_{\text{CDW}}$ \cite{salvo1976}. $1T$-TiSe$_2$ is also superconducting upon Cu intercalation \cite{Morosan2006a}, and under pressure \cite{Kusmartseva2009a}, thus offering the way for the study of the interplay between ordered CDW phase and the superconducting state \cite{Joe2014}. 

However, the origin of the CDW transition itself is not yet unambiguously determined. Although many propositions have been made since the 70's, there still remain two competing hypotheses for the mechanism driving the CDW formation. Several studies have highlighted strong electron-phonon \cite{yoshida1980, Holt2001a, Weber2011b} coupling suggesting a Jahn-Teller (JT) mechanism of the CDW transition (first order JT \cite{Hughes1977} or pseudo-JT(PJT) \cite{whangbo1992}). Electron-hole correlations leading to the excitonic insulator \cite{Wilson1978a, Cercellier2007a} or inducing fluctuations responsible for phonon softening \cite{Monney2012, Monney2015} have been also invoked as well as a cooperative combination of both effects \cite{VanWezel2010c, Kidd2002}. 

Very recently, significant insights favoring the JT scenario has been obtained by using ultra-broadband terahertz pulses to separately trace the coexisting lattice and electronic orders \cite{Porer2014}. Taking profit of the selectivity in the dynamics of elementary electronic and structural processes \cite{Hellmann2012, Rohwer2011}, it has been demonstrated that JT-like CDW can exist in a metastable non-thermal phase without excitonic correlations. However, time-resolved experiments \cite{Porer2014, Hellmann2012, Rohwer2011, Mohr-Vorobeva2011}, probe the CDW out-of-equilibrium and thus cannot be straightforwardly extended to the thermal equilibrium states. Also, most of the studies devoted to this controversy used experimental techniques non-local in essence and therefore not probing directly local, real-space characteristics of the CDW. 

\begin{figure*}[t]
\includegraphics[scale=1]{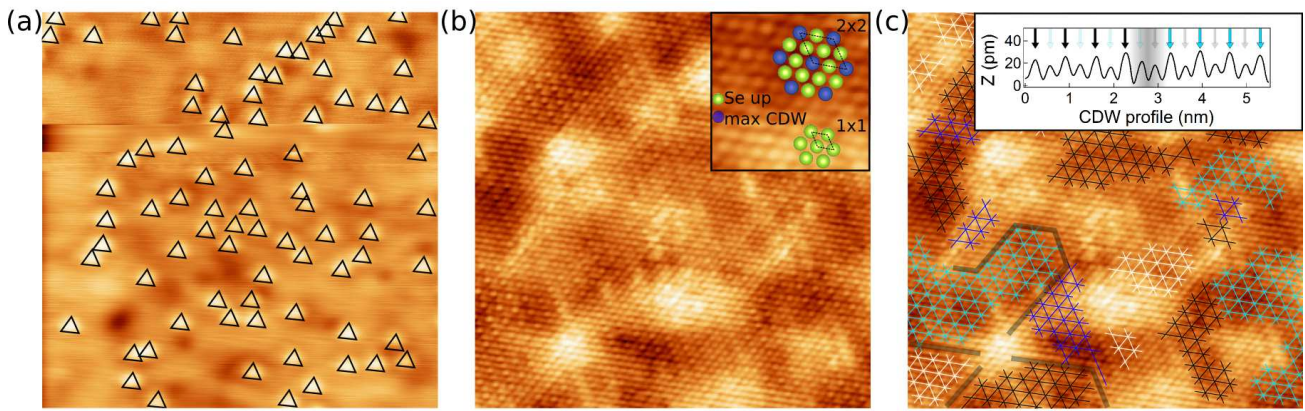}
\caption{\label{fig1} (Color online) (a) 17$\times$17nm$^2$ constant current STM images of the 900\celsius$~$-grown 1$T$-TiSe$_2$ crystal showing the intercalated-Ti defects (black triangles), V$_{\text{bias}}=-1$V, I$=0.2$nA. (b) Constant current STM image at $-0.1$V bias voltage and $0.2$ nA current set point of the same region as (a) showing the CDW charge modulation and its perturbation by the presence of Ti defects. The inset presents an eye guide of the 2x2 and 1x1 charge modulations. Green spheres correspond to Se atoms of the topmost layer (Se up) and the blue ones to the maxima of the 2x2 charge modulation (max CDW). (c) Same STM image as (b) with added meshes highlighting the different CDW domains. Four colors have been used to differentiate between the four possible kinds of modulations that are phase shifted to each other. The gray lines show examples where CDW phase-slips occur. The inset shows a height profile through an atomically abrupt CDW phase-slip.} 
\end{figure*}

Using scanning tunneling microscopy (STM) it is possible to gain insight on the impact of defects on the local CDW properties \cite{Chatterjee2015, Weitering1999, Wu1989}. In 1$T$-TiSe$_2$, Ti self-doping is known to occur depending on the crystal growth temperature \cite{salvo1976}, with a dramatic effect on resistivity. Intercalated-Ti atoms lead to electron-donor impurity states close to the Fermi energy \cite{Hildebrand2014}, enhance the Coulomb screening and tend to reduce electron-hole correlations. Therefore, the real-space STM investigation of the CDW in Ti self-doped crystals can in principle reveal the microscopic nature of the phase transition.

The present STM study of Ti self-doped 1$T$-TiSe$_2$ shows that accumulation of intercalated-Ti atoms act as preferential nucleation centers for the CDW. The CDW breaks up in randomly phase-shifted nanodomains with subsisting commensurate 2x2 charge modulation and separated by atomically sharp phase slips. 
We find a close relationship between the CDW domain size and the Ti-doping density and demonstrate that the nucleation-growth mechanism lying behind the CDW formation is unique, regardless of the intercalated-Ti density. 
Our observations together with density functional theory (DFT) calculations support a local origin of the CDW as driven by a PJT instability, not necessitating long-range correlations. We interpret the dramatic influence of Ti-doping on the anomalous resistivity peak as a confinement effect of excitons within CDW domains.

The 1$T$-TiSe$_2$ single crystals were grown at 770 \celsius$~$, 860 \celsius$~$ and 900 \celsius~ by iodine vapor transport, therefore containing increasing concentrations of Ti doping atoms \cite{salvo1976}. Resistivity measurements were performed by a standard four-probe method using a lock-in as current source and voltage meter. The samples were cleaved \textit{in-situ} below 10$^{-7}$mbar at room temperature. Constant current STM images were recorded at 4.7 K using an Omicron LT-STM, with bias voltage V$_{\text{bias}}$ applied to the sample. Base pressure was better than 5$\times$10$^{-11}$mbar.

DFT model calculations were performed using the plane-wave pseudopotential code VASP \cite{Kresse1993, Kresse1996}, version 5.3.3. Projector augmented waves \cite{Kresse1999} were used with the Perdew-Burke-Ernzerhof (PBE) \cite{Perdew1996} exchange correlation functional. The cell size of our model was 28.035 \AA~$\times$ 28.035 \AA. The 1$T$-TiSe$_2$ surface was modeled with two layers and the bottom Se layer fixed. A Monkhorst-Pack mesh with 2$\times$2$\times$1 $k$ points was used to sample the Brillouin zone of the cell. The parameters gave an energy difference convergence of better than 0.01 eV. During structural relaxations, a tolerance of 0.03 eV/\AA~ was applied.

Figure \ref{fig1}(a) presents filled-state STM images recorded at -1 V of the 1$T$-TiSe$_2$ surface of a high Ti-doped 1$T$-TiSe$_2$ crystal grown at 900\celsius$~$. Bright defects are clearly seen and correspond to Ti atoms intercalated in the vdW gap of the pristine $1T$-TiSe$_2$ atomic structure \cite{Hildebrand2014}. 
The defect density is 2.57$\pm$0.22$\%$ as extracted from statistics made on large STM images and similar to bulk measurements by Di Salvo \textit{et al.} on a 900\celsius$~$-grown crystal \cite{salvo1976}. Thus, even if the densities of Ti defects estimated from our STM images mainly account for Ti intercalation in the first vdW gap, they are representative of the \textit{bulk} Ti-doping density. 

In a recent study, it has been shown that the defect density associated with low-doped $1T$-TiSe$_2$ samples does not affect the \textit{long-range} CDW order \cite{novello2015}. For the highly doped crystal and looking at regions with high density of intercalated Ti atoms, atomically-resolved STM images (representing the electronic density) near the Fermi energy [Fig. \ref{fig1}(b)] show that the CDW is locally completely absent (bright regions). Approximately 50$\%$ of the $1T$-TiSe$_2$ surface clearly exhibits \textit{short-range} CDW charge modulation following the well-known 2x2 commensurate pattern [see inset Fig. \ref{fig1}(b)] in regions associated with low defect density. 

Locating the CDW maxima with respect to the underlying Se atomic layer [Fig. \ref{fig1}(c)] now reveals a CDW patterning of nanometer size. Several CDW patterns which differ in their phase relationship with the $1T$-TiSe$_2$ lattice are observed and correspond to the four possible configurations of a 2x2 charge modulation. Interestingly, the CDW boundaries in surface areas free of accumulation of intercalated-Ti consist of atomically-sharp phase slips between adjacent domains [see inset Fig. \ref{fig1}(c)] and appear to occur at random positions of the surface with the charge modulation on either side of the phase slip staying commensurate with the underlying lattice. Isolated Ti impurities are mainly observed within domains suggesting that the CDWs emerged around the bright regions [see bottom, left of Fig. \ref{fig1}(c)] associated with an accumulation of intercalated-Ti defects. 

Therefore, our results strongly indicate that the CDW domains patterning originates in a CDW nucleation-growth process from intercalated-Ti agglomerates. Upon temperature lowering, the nucleating CDWs grow and meet at phase-slip boundaries whose sharpness reflects the very short CDW coherence length \cite{Rossnagel2011, Holt2001a}. 

At first sight, it may be surprising that a commensurate CDW can break into short-range phase coherent domains in the presence of impurities. Indeed, as previously pointed out by Wu and Lieber who observed commensurate CDW domains on another system (Ti-doped TaSe$_2$) \cite{Wu1990}, a commensurate CDW phase that is already strongly coupled to the crystal lattice tempts to remain unperturbed by impurity pinning according to a seminal work of McMillan \cite{McMillan1975}. However, in contrast to McMillan's Landau theory calculations that describe how a pre-existing charge modulation reacts to the random field introduced by impurities \cite{McMillan1975}, we are concerned here by the emergence of the CDW in with defects already present. Then, the size and arrangement of the CDW domains as well as the domain boundary locations are fully determined by the density of intercalated-Ti nucleation centers, their distribution and the cooling kinetics. 

\begin{figure}[t]
\includegraphics[scale=1]{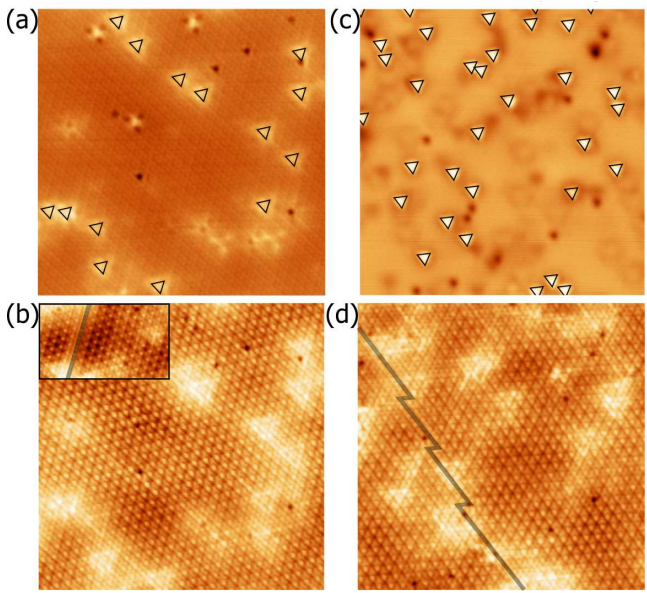}
\caption{\label{fig2} (Color online) (a) 17$\times$17nm$^2$ constant current STM image of the 770\celsius$~$-grown 1$T$-TiSe$_2$ crystal showing the intercalated-Ti defects (black triangles), V$_{\text{bias}}=+1$V, I$=0.2$nA. (b) Constant current STM image at $-0.1$V bias voltage and $0.2$ nA current set point of the same region as (a). (c), (d) same as (a), (b) for the 860\celsius$~$-grown 1$T$-TiSe$_2$ sample. V$_{\text{bias}}=-1$V for (c). The gray lines on (b) and (d) show surface regions where a CDW phase-slip occurs. The inset in (b) is a 7.8$\times$4.5nm$^2$ constant current STM image showing the presence of a CDW phase-slip on the 770\celsius$~$-grown sample.}
\end{figure}

STM measurements carried on $1T$-TiSe$_2$ crystals grown at lower temperature further support our CDW nucleation-growth scenario. Figure \ref{fig2} shows 17$\times$17nm$^2$ constant current STM images of a 770\celsius$~$ [Fig. \ref{fig2}(a), (b)] and 860\celsius$~$-grown [Fig. \ref{fig2}(c), (d)] 1$T$-TiSe$_2$ crystals. Comparing both cystals with the 900\celsius$~$-grown one for V$_{\text{bias}}=\pm{1}$V, we clearly see that the density of randomly distributed Ti defects [highlighted with triangles on Fig. \ref{fig1}(a) and Fig. \ref{fig2}(a), (c)] increases with growth temperature. From statistics made on large STM images (not shown), we obtain 0.75$\pm$0.10$\%$ and 1.21$\pm$0.14$\%$ for the crystals grown at 770\celsius$~$ and 860\celsius$~$, respectively compared to the 2.57$\pm$0.22$\%$ of the 900\celsius$~$-grown $1T$-TiSe$_2$.

Identically to the highly-doped crystal, STM images near Fermi level [Fig. \ref{fig2}(b), (d)] now reveal for both crystals the presence of atomically-sharp phase slips of the CDW modulation. Whereas for the lowest doped crystal 60$\times$60nm$^2$ images are needed to observe a shift of the CDW phase [inset [Fig. \ref{fig2}(b)], every 17$\times$17nm$^2$ images of the 860\celsius$~$-grown crystal show at least one phase slip. This not only demonstrates the close relationship between the CDW domain sizes and the Ti-doping density but especially reveals that the nucleation-growth mechanism lying behind the CDW formation in the three Ti-doped crystals is the same.  

\begin{table}[b]
\caption{\label{tab1}DFT-calculated structural parameters of 1$T$-TiSe$_2$ with and without intercalated Ti atom in the vdW gap. Two TiSe$_6$ octahedra separated by one vdW gap are considered: Interatomic distances between Se atoms of the topmost layer ($d_{Se-Se}$), between the Se neighboring the intercalated-Ti defect ($d_{Se-Se}$(vdW)), between Ti atoms along the $c$-axis ($d_{Ti-Ti}$), and between Ti and Se atoms of the TiSe$_6$ octahedra ($d_{Ti-Se}$).}
\begin{ruledtabular}
\begin{tabular}{lcc}
& Ti-doped 1$T$-TiSe$_2$ & 1$T$-TiSe$_2$\\ 
& (\AA) & (\AA)\\
\hline
$d_{Se-Se}$ & 3.50 & 3.54 \\
$d_{Se-Se}$(vdW) & 3.64 & 3.54 \\
$d_{Ti-Ti}$ & 6.04 & 6.31\\
$d_{Ti-Se}$ & 2.58 & 2.56\\
\end{tabular}
\end{ruledtabular}
\end{table}

At this stage, we have shown that the CDW can well exist on a very small area and that its pinning to the nucleation centers is energetically more favorable than maximizing a domain size or avoiding phase-slips. Also, the coherence length of the CDW is definitely small suggesting that the strong-coupling theory and a local chemical bonding picture apply \cite{Rossnagel2011}. This naturally indicates a local origin of the CDW as driven by a PJT instability \cite{whangbo1992, bersuker2006}, i.e., not necessitating long-range electronic correlations.   

\begin{figure}[t]
\includegraphics[scale=1]{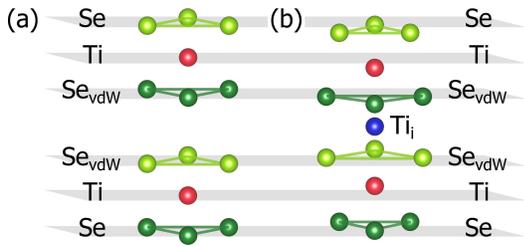}
\caption{\label{fig3} (Color online) Structural models of 1$T$-TiSe$_2$ as obtained by DFT. Two TiSe$_6$ octahedra separated by one vdW gap are considered (a) Pristine 1$T$-TiSe$_2$ (b) Ti-doped 1$T$-TiSe$_2$. The amplitude of displacements induced by Ti intercalation (Ti$_i$) have been enhanced.}
\end{figure}

Let us now consider the impact of the intercalation of Ti on the 1$T$-TiSe$_2$ atomic structure. Table \ref{tab1} presents DFT-calculated, relaxed interatomic distances of the normal phase of 1$T$-TiSe$_2$ with and without one intercalated-Ti atom in the vdW gap. The impact of the intercalation is rather strong and of the order of magnitude of the PLD. As exemplified in Fig. \ref{fig3}, the major effect of the intercalation is the strong decrease (4.3\%) of the distance between Ti atoms along the $c$-axis ($d_{Ti-Ti}$) and the symmetric displacement of the Se atoms surrounding the intercalated-Ti defect. Their in-plane distance ($d_{Se-Se}$(vdW)) increases by 2.7 \% whereas the distance between Se atoms of the topmost layer ($d_{Se-Se}$) decreases by 1.2 \% overall leading to elongated (0.8 \%) Ti-Se bonds ($d_{Ti-Se}$) of the TiSe$_6$ octahedra.

Therefore, the intercalation of one Ti atom in the vdW gap does not favor the CDW instability since it acts against the PLD where 3/4 of the Ti-Se bonds are shortened \cite{salvo1976}. This explains why one isolated Ti defect does not act as a nucleation center for the CDW and also why the CDW charge modulation is highly perturbed just above the intercalated-Ti atom. However, at intercalated-Ti agglomerate edges, the coherent superposition of symmetric Se displacements are expected to induce shortening of the Ti-Se bonds favorable for driving the nucleation of multiple phase-shifted domains in a PJT scenario \cite{whangbo1992}. 

Finally our findings provide insight on the anomalous resistivity peak known to appear around $T_{\text{CDW}}$. Because of a favorable band configuration with energetically close positioned electron- and hole-like bands, it has been interpreted as a reduction of the density of free charge carriers due to electron-hole pairing \cite{Wilson1978a, Cercellier2007a}. Its disappearance at lower temperatures is then related to the development of the full CDW gap with unfavorable conditions for the excitonic pairing. 

\begin{figure}[t]
\includegraphics[scale=1]{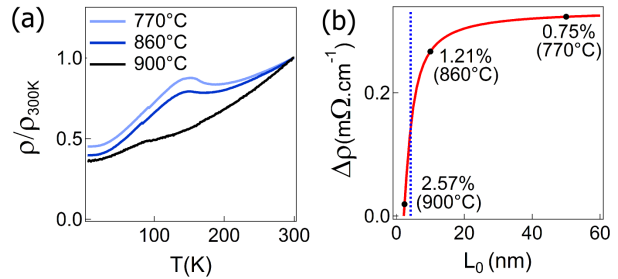}
\caption{\label{fig4} (Color online) (a) Temperature dependent resistivity curves measured on 770\celsius$~$ (blue curve), 860\celsius$~$ (dark blue curve), and 900\celsius$~$(black curve) grown 1$T$-TiSe$_2$ crystals. The curves are normalized to the resistivity measured at room temperature. b) Evolution of the intensity of the resistivity peak as a function of the CDW domains size. The dashed blue curve indicates the typical value of the effective Bohr diameter of excitons in 1$T$-TiSe$_2$.}
\end{figure}

Figure \ref{fig4}(a) shows the relative resistivity $\rho/\rho_{300K}$ of the three measured crystals. A maximum is clearly observed for the 770\celsius$~$ and 860\celsius$~$-grown crystals whereas it is almost absent for 900\celsius. 
In the evolution of the anomalous resistivity peak amplitudes \footnote{The anomalous resistivity peak amplitude was obtained from the temperature dependent resistivity curves after extraction of the semi-metallic background using an experimental curve of highly Cu-doped 1$T$-TiSe$_2$ that does not show the anomalous resistivity peak.}, as a function of the average domain sizes $L_0$ extracted from our STM measurements \footnote{$L_0$ was estimated from STM images by measuring the average distance between boundaries of CDW domains along the three crystal directions.} plotted in [Fig. \ref{fig4}(b)] there is a dramatic change between the sample grown at 860\celsius$~$ and 900\celsius$~$ despite the only slight diminution of the domain size from 10 nm to 2.5 nm. It is also interesting to note in Fig. \ref{fig4}(b) that a small doping increase from 0.75 \% (770\celsius) to 1.21 \% (860\celsius) is accompanied by a small change in the resistivity but induces a huge reduction of the mean domain size $L_0$.

The abrupt but continuous change of the resistivity maximum with doping can now be understood in terms of exciton confinement within the CDW nanodomains, as discussed for Wannier-Mott excitons in three-dimensional quantum wells \cite{Kayanuma1986}. As the typical size of the CDW domain is reduced to 1.5-2 times the effective Bohr diameter of the bulk exciton ($d_{exc.}=$4.2 nm \cite{Pillo2000}) [dashed-blue curve Fig. \ref{fig4}(b)], the electron-hole pairs initially confined as quasiparticles start to break up into individual charge carriers \cite{Kayanuma1986}. The result is the disappearance of the exciton-related resistivity maximum for sufficient doping or sufficiently small CDW domains.

The resistivity peak attributed to excitons thus reflects sufficiently long-range electronic correlations in the CDW state. In that sense, the excitonic pairing is an epiphenomenon of the CDW phase transition driven by electron-phonon interaction that allows for the direct coupling of electron and hole bands. Exciton-phonon coupling may thus simply reinforce the CDW state \cite{Phan2013}, without playing the predominant role in \textit{driving} the phase transition itself.

To summarize, we report the first observation of short-range phase coherent CDW nano domains in self-doped 1$T$-TiSe$_2$. Our results support a PJT instability as the origin of the CDW patterning. Evidence for confinement effects of electron-hole pairs within CDW domains further provide a coherent picture of the $1T$-TiSe$_2$ CDW phase transition and its relation to excitons.

\begin{acknowledgments}
This project was supported by the Fonds National Suisse pour la Recherche Scientifique through Div. II. We would like to thank C. Monney, C. Renner, and A. M. Novello for motivating discussions. Skillful technical assistance was provided by F. Bourqui, B. Hediger and O. Raetzo.

B.H. and T.J. contributed equally to this work.
\end{acknowledgments}

\bibliography{library1}
\end{document}